\documentclass[sigconf, nonacm]{acmart}

\usepackage{algorithm, setspace}
\usepackage{comment}
\usepackage{mathtools}
\usepackage{footnote}
\usepackage{tabularx}
\usepackage{booktabs}
\usepackage[noend]{algpseudocode}
\usepackage{multicol}
\usepackage{multirow}
\usepackage{subcaption}
\usepackage{cleveref}
\usepackage{xcolor}
\floatname{algorithm}{Algorithm}

\renewcommand{\arraystretch}{0.75}

\usepackage{tcolorbox}
\usepackage{verbatim}
\usepackage{array}
\newcolumntype{H}{>{\setbox0=\hbox\bgroup}c<{\egroup}}

\def\ours{{Meta Engine}}

\newcommand\vldbdoi{XX.XX/XXX.XX}
\newcommand\vldbpages{XXX-XXX}
\newcommand\vldbvolume{14}
\newcommand\vldbissue{1}
\newcommand\vldbyear{2020}
\newcommand\vldbauthors{\authors}
\newcommand\vldbtitle{\shorttitle} 
\newcommand\vldbavailabilityurl{URL_TO_YOUR_ARTIFACTS}
\newcommand\vldbpagestyle{plain}

\begin{document}

\title{Beyond Single-Modal Analytics: A Framework for Integrating Heterogeneous LLM-Based Query Systems for Multi-Modal Data}

\author{*Ruyu Li}
\email{liruyu@hawaii.edu}
\affiliation{%
  \institution{University of Hawaii at Manoa}
}

\author{*Tinghui Zhang}
\email{tinghui.zhang@ufl.edu}
\affiliation{%
  \institution{University of Florida}
}

\author{Haodi Ma}
\email{ma.haodi@ufl.edu}
\affiliation{%
  \institution{University of Florida}
}

\author{Daisy Zhe Wang}
\email{daisyw@ufl.edu}
\affiliation{%
  \institution{University of Florida}
}

\author{Yifan Wang}
\email{yifanw@hawaii.edu}
\affiliation{%
  \institution{University of Hawaii at Manoa}
}

\begin{abstract}
Many LLM-based semantic query systems have been proposed to support semantic querying over structured and unstructured data. 
However, this rapid growth has produced a fragmented ecosystem. Applications face significant integration challenges due to (1) disparate APIs of different semantic query systems and (2) a fundamental trade-off between specialization and generality. Many semantic query systems are highly specialized (e.g., DocETL for long documents, StructGPT for table QA), offering state-of-the-art performance within a specific task or a single modality but struggling with multi-task or multi-modal data. Conversely, some "all-in-one" systems (e.g., LOTUS) handle multiple modalities but often exhibit suboptimal performance compared to their specialized counterparts in specific modalities. In short, the existing systems still have significant limitations when applied to comprehensive tasks over complex multi-modal data.  

This paper introduces \ours, a novel framework for building "query system on query systems", designed to resolve those aforementioned challenges. Meta Engine is a unified framework to integrate heterogeneous, specialized LLM-based query systems into one query processing pipeline, such that the specialities of different systems are combined to handle complex scenarios. It also provides unified interfaces for users to query through the pipeline, keeping the disparate APIs of underlying systems transparent to users. \ours{} integrates several key components for APIs, query preprocessing, query planning and execution, as well as postprocessing, which cover the whole semantic querying lifecycle. The components are extensible, flexible, and composable. Users can extend or replace any of them with their own implementation, and flexibly choose components to customize their pipelines.       
This design makes \ours{} easy to use and able to combine the distinct strengths of different semantic query systems, achieving high-performance multi-modal semantic processing from a unified entry point. In the evaluation, Meta Engine consistently outperforms all baselines, yielding 3–6x higher F1 in most cases and up to ~24x on specific datasets. 
\end{abstract}

\maketitle

\pagestyle{\vldbpagestyle}
\begingroup\small\noindent\raggedright\textbf{PVLDB Reference Format:}\\
\vldbauthors. \vldbtitle. PVLDB, \vldbvolume(\vldbissue): \vldbpages, \vldbyear.\\
\href{https://doi.org/\vldbdoi}{doi:\vldbdoi}
\endgroup
\begingroup
\renewcommand\thefootnote{}\footnote{\noindent
This work is licensed under the Creative Commons BY-NC-ND 4.0 International License. Visit \url{https://creativecommons.org/licenses/by-nc-nd/4.0/} to view a copy of this license. For any use beyond those covered by this license, obtain permission by emailing \href{mailto:info@vldb.org}{info@vldb.org}. Copyright is held by the owner/author(s). Publication rights licensed to the VLDB Endowment. \\
\raggedright Proceedings of the VLDB Endowment, Vol. \vldbvolume, No. \vldbissue\ %
ISSN 2150-8097. \\
\href{https://doi.org/\vldbdoi}{doi:\vldbdoi} \\
}\addtocounter{footnote}{-1}\endgroup
\def\thefootnote{*}\footnotetext{These authors contributed equally to this work.}\def\thefootnote{\arabic{footnote}}
\ifdefempty{\vldbavailabilityurl}{}{
\vspace{.3cm}
\begingroup\small\noindent\raggedright\textbf{PVLDB Artifact Availability:}\\
The source code, data, and/or other artifacts have been made available at \url{\vldbavailabilityurl}.
\endgroup
}

\section{Introduction}
\label{sec:intro}
Semantic query has become more and more demanded in data management systems, which is an important way to access and analyze unstructured data from various modalities. Comparing to traditional SQL queries that use standard and structured syntax to query the data, semantic queries usually rely on natural language (NL) to semantically process unstructured data like image, text, etc.      
Therefore, semantic query processing requires the capabilities of understanding NL request and capturing the semantics in data, which is hard in the past years and the solutions are usually limited~\cite{sem-web-lu2002semantic, sem-web-spanos2012bringing, vec-in-db-1, vec-in-db-10184805, vec-in-db-bandyopadhyay2020drugdbembedsemanticqueries}.
The advance of Large Language Models (LLMs) leads to a paradigm shift in this, resulting in the development of advanced semantic query systems~\cite{vec-in-db-1, vec-in-db-10184805, vec-in-db-bandyopadhyay2020drugdbembedsemanticqueries,lotus-patel2024lotusenablingsemanticqueries,docetl,structgpt-jiang2023structgptgeneralframeworklarge,dspy-khattab2023dspycompilingdeclarativelanguage, zhang2025scopegenerativeapproachllm}. They enable applications to semantically reason and analyze over complex (like multi-modal) datasets using natural language. 


Despite this rapid progress, the current ecosystem of LLM-based query systems is highly fragmented, presenting critical challenges that hinder the comprehensive use of them on complex tasks.
First, diverse systems result in disparate and incompatible APIs. 
Consequently, developers seeking to build applications that leverage the strengths of multiple systems are forced to write complex integration logic, increasing both development and maintenance overhead.
Second challenge is the crucial trade-off between specialization and generality. Many state-of-the-art semantic query systems are highly specialized, achieving exceptional performance by optimizing for specific tasks in single modality. For instance, DocETL \cite{docetl} excels at processing large-scale long text documents, while StructGPT \cite{structgpt-jiang2023structgptgeneralframeworklarge} is designed specifically for semantic querying of tabular data. Their high specialization in specific domain/modality prevents them from integrating and processing information across different modalities. Conversely, "all-in-one" systems like LOTUS \cite{lotus-patel2024lotusenablingsemanticqueries} attempt to provide a unified solution for multi-modal data (e.g., table, text, and image), but the cost of such a generalization is often suboptimal performance on specific single modality, like the image QA quality of LOTUS is usually lower than a pure VLM in our evaluation. 

These challenges raise a critical need for a unified framework that can (1) comprehensively leverage the strengths of different specialized systems to achieve high-performance and truly multi-modal semantic query processing, and (2) provide a unified and easy-to-use interface for user interaction.

\begin{figure}[ht]
  \centering
  \includegraphics[width=1.05\columnwidth]{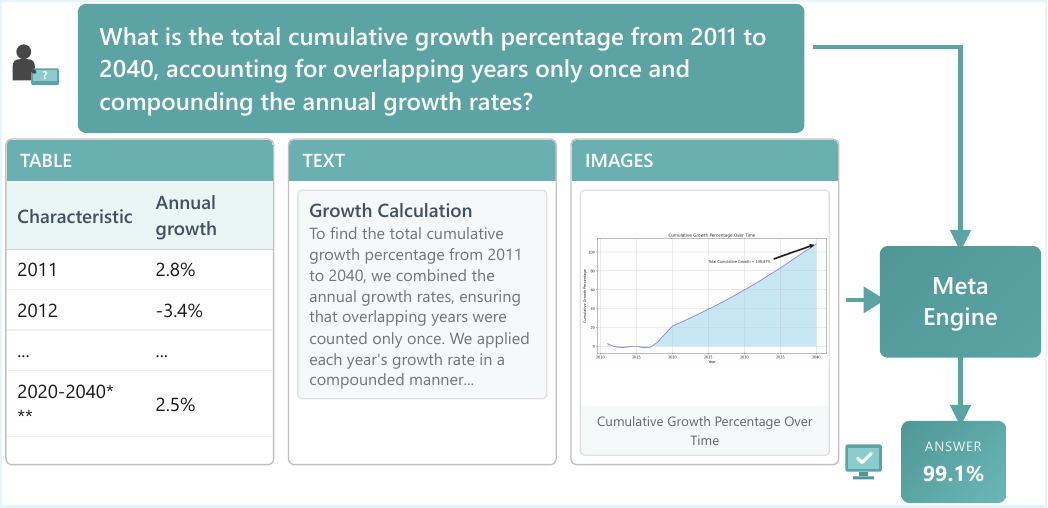}
  \caption{An End-to-End Multimodal Analytical Query Example in Meta Engine}
  \label{fig:exp-q}
  \vspace{-10pt}
\end{figure}

In this paper, we propose \ours{}, a novel framework to build the "query system over query systems". \ours{} functions as an comprehensive framework with flexible, extensible and composable components, designed for integrating heterogeneous LLM-based semantic query systems into a cohesive pipeline. By adaptively scheduling the different systems, it can intelligently combine their specialities to achieve a significant performance gain on complex analytics. Additionally, to address the disparate API issue, \ours{} provides two sets of unified interfaces: a natural language interface and a set of functions including 
data operators and building block operators. With the interfaces, users can customize the pipeline flexibly, and query the pipeline using either NL questions or programmatically calling the specific operators.      
These interfaces keep the disparate APIs of underlying systems transparent to users, as well as ensure an easy pipeline construction.   
Furthermore, \ours{} supports pipeline construction for not only typical question answering, but also broader data analytic tasks on multi-modal data. 
For example, in one of our evaluation dataset ~\cite{rahman2025text2vischallengingdiversebenchmark}, each natural language query specifies an analytical task over structured tables, text and images. An example query and supporting documents are shown in Figure~\ref{fig:exp-q}.  

\ours{} includes several major components: 
(1) APIs: The APIs include an NL query interface and several functions for the data operators and the building blocks of the pipeline. Table~\ref{tab:api} lists the major APIs. Specifically, the building blocks are those used to build the pipeline, while the data operators are those doing the analytics over data in runtime.
(2) Query preprocessing: We design the Query Decomposer that smartly breaks down complex NL queries into simpler sub-queries. This component accepts the user's NL queries and intelligently decompose each query to several sub-queries. 
A dependency graph is used to manage the dependencies between subqueries for each query. 
(3) Query planning:
After decomposition, \ours{} maps each subquery to an executable data operator. Since each sub-query is simple enough, one analytic data operator can process it. Instead of selecting a single operator, \ours{} will select multiple candidate data operators ranked by the selection confidence, where higher-ranked operators are more likely to correctly process the sub-query. More details are in Section~\ref{sec:building-blocks} and \ref{sec:op-selector}. 
(4) Query execution:
Query execution includes execution of data operators and subqueries. Essentially, executing a subquery is to execute its corresponding data operators, with additional processing to the subquery afterwards.  
In data operator level, an operator will be dispatched to the most appropriate specialized system (adapter) by a Query Router to execute. 
In subquery level, for each subquery, its ranked candidate operators (from query planning) will be executed one by one as above, following the confidence order (from high to low), until any one successfully processes the subquery and returns valid result. 
Furthermore, in query level, we propose two optimizations: iterative sub-query refinement and dependency-based subquery parallelism for the execution. More details are in Section~\ref{sec:op-selector}.  
(5) Adapters: As an important part of query execution, adapters serve as wrappers for the underlying LLM query systems to abstract away the heterogeneity of those systems. Each adapter manages the unique interfaces of a system and exposes standardized interfaces to \ours{} users. 
(6) Query postprocessing: After all subqueries are executed, \ours{} provides an optional Result Aggregator to comprehend the results of sub-queries into the coherent final result, with necessary refinement like deduplicating. If users tend not to do the postprocessing, the final result will be the result of the last subquery.

Additionally, \ours{} is extensible that allows users to extend most components, e.g., users can replace the decomposer with their own implementation. \ours{} is also composable and flexible, as most components are optional in a pipeline and users are free to build the pipeline as they like. For example, users can pass the original query to $select\_op$ without decomposition, or directly call specific data operator to process the query/subquery while skipping $select\_op$.    
With the composable, extensible components and unified APIs in \ours{}, users can easily customize a pipeline that integrates different specialized systems for complex multi-modal analytics. The architecture and optimization in \ours{} resolves the specialist-generalist trade-off, enabling applications to benefit from high-performance, multi-modal processing with a unified solution. To our best knowledge, \ours{} is among the first works to build such a unified and flexible framework for integrating heterogeneous LLM systems in one pipeline.  

The main contributions of this work are as follows:
\begin{enumerate}
    \item We propose Meta Engine, a novel unified framework to efficiently build multi-modal query processing pipeline that integrates different specialized LLM query systems.
    \item We develop composable, extensible and flexible architecture and components, with novel optimization techniques. 
    \item We design unified APIs to unify the heterogeneous interfaces of different query systems, significantly simplifying pipeline development.
    \item We conduct extensive evaluation that shows the effectiveness of our proposed system, as well as ablation studies to show the impacts of each major component.  
\end{enumerate}

The remainder of this paper is structured as follows: Section~\ref{sec:related-work} introduces prior works related to \ours{}. Section~\ref{sec:arch} and 4 introduce the architecture, workflow and technical details of our method. Section~\ref{sec:exps} reports and analyzes experiment results.

\vspace{-8pt}
\begin{table*}[htbp]
    \centering
    \caption{Summary of Key APIs: (1) End-to-end NL interface, (2) building block APIs for pipeline customization, and (3) data operator APIs for multi-modal analytics.}
    \vspace{-10pt}
    \label{tab:api}
    \renewcommand{\arraystretch}{1.25}
    \footnotesize
    \setlength{\tabcolsep}{3.5pt}
    \newcolumntype{Y}{>{\hsize=1.4\hsize}X}
    \newcolumntype{Z}{>{\hsize=0.6\hsize}X}
    \begin{tabularx}{\textwidth}{@{} l l Y Z @{}}
        \toprule
        \textbf{API} & \textbf{Return} & \textbf{Description} & \textbf{Algorithm} \\
        \midrule
        \multicolumn{4}{@{}l}{\textit{\textbf{End-to-End Interface}}} \\[1pt]
        \texttt{query}($q$) &
        Query result &
        End-to-end pipeline execution for NL query $q$. &
        --- \\
        \midrule
        \multicolumn{4}{@{}l}{\textit{\textbf{Building Block APIs}}} \\[1pt]
        \texttt{decompose}($q$, $check$, $max\_out$) &
        Sub-queries &
        Decomposes $q$ into sub-queries; conditional on complexity if $check{=}\text{True}$. &
        Dynamic decomposition \\[2pt]
        \texttt{select\_op}($q$, $ops$) &
        Ranked op.\ list &
        Ranks candidate operators for $q$ by confidence. &
        LM-based classification \\[2pt]
        \texttt{exec\_op}($ops$, $r$:\{S$|$L\}) &
        Op.\ results &
        Routes operators to adapters and executes; $r$ selects routing strategy. &
        History / trained routing \\[2pt]
        \texttt{exec\_subq}($qs$, $par$) &
        Subquery results &
        Executes sub-queries via \texttt{select\_op} and \texttt{exec\_op}; $par$ enables parallelism. &
        Iterative refinement + parallel exec. \\[2pt]
        \texttt{route}($ops$, $adpts$, $r$:\{S$|$L\}) &
        Adapter assign. &
        Maps each operator to its best adapter; $r$ selects routing strategy. &
        Statistics / learning-based pred. \\[2pt]
        \texttt{agg\_results}($res$, $q$) &
        Final answer &
        Aggregates sub-query results into the final answer for $q$. &
        LLM summarization \\
        \midrule
        \multicolumn{4}{@{}l}{\textit{\textbf{Data Operator APIs}}} \\[1pt]
        \texttt{TextAnalytics}($q$) &
        Text analytics result &
        Executes analytics specified by $q$ on text documents. &
        Adapter-based (text) \\[2pt]
        \texttt{ImageAnalytics}($q$) &
        Image analytics result &
        Executes analytics specified by $q$ on image documents. &
        Adapter-based (image) \\[2pt]
        \texttt{TableAnalytics}($q$) &
        Table analytics result &
        Executes analytics specified by $q$ on tables. &
        Adapter-based (table) \\
        \bottomrule
    \end{tabularx}
    \vspace{-8pt}
\end{table*}

\section{Related work}
\label{sec:related-work}
The study of semantic querying has historically aimed to bridge the gap between rigid data schemas and human-like understanding. In the pre-LLM era, visions like the Semantic Web~\cite{sem-web-lu2002semantic, sem-web-spanos2012bringing} embeds data semantics directly into the web via ontologies and knowledge graphs, but such approaches struggle with the high overhead of maintaining global ontologies. 
Subsequent approaches integrated advanced deep learning methods with databases. For example, \cite{vec-in-db-1, vec-in-db-10184805, vec-in-db-bandyopadhyay2020drugdbembedsemanticqueries} utilize deep-learning-generated embeddings (e.g., word2vec~\cite{w2v-mikolov2013efficientestimationwordrepresentations}) and User-defined function (UDF) to compute semantic similarities between values or columns. However, these systems often lack a more comprehensive view of inter-column relationships and required users to manage complex UDFs. Recent work leverages large language models to perform high-level planning and optimization for semantic queries, automatically reasoning over query structure and execution strategies across complex data workflows~\cite{Wang_2025}. Furthermore, embedding is good at measuring the semantic \textit{similarity} between objects, but cannot capture more complex semantics. For example, based on the embedding distance, users can determine how similar two values/objects are, but hard to know the relationship between them. 

The advent of Large Language Models (LLMs) marked a paradigm shift, enabling systems to interpret flexible natural language predicates and execute complex reasoning without model modifications or extensive tuning.
Recent semantic query systems focus on deeply integrating LLMs into data processing pipelines to enhance accuracy, usability and flexibility, benefiting from the strong semantic representation capabilities of large-scale language models across downstream ~\cite{Li_2025}. They usually define declarative programming interfaces (\textit{semantic operators}) that extends the relational model with composable AI-based operators, powered by LLM. The operators will then be composed into complex semantic queries. 
For example, StructGPT~\cite{structgpt-jiang2023structgptgeneralframeworklarge} represents a "reading-then-reasoning" framework with operators like \textit{Extract\_SubTable} and \textit{Extract\_Neighbor\_Relations} to process structured data using LLM. 
LOTUS~\cite{lotus-patel2024lotusenablingsemanticqueries} integrates semantic operators like semantic join and filter into Python Pandas Dataframe API, with performance optimization for each operator. 
UQE~\cite{uqe-dai2024uqequeryengineunstructured} integrates semantic operators into SQL in relational database environment, treating unstructured text as "virtual columns" to allow seamless semantic queries in SQL. 
DocETL~\cite{docetl} extends these capabilities to agentic document processing, designing LLM-powered operators to construct complex transformation pipelines for long-form documents.
DSPy~\cite{dspy-khattab2023dspycompilingdeclarativelanguage} defines its declarative programming paradigm with a focus on prompt engineering, allowing developers to define high-level modules that the framework automatically optimizes the task pipeline for better prompts and weights. 
But those semantic query systems are either highly specialized (like DocETL focuses on long document processing) or general-purposed with average performance on most tasks.    




\section{Architecture}
\label{sec:arch}

\begin{figure}[ht]
  \centering
  \includegraphics[width=1\columnwidth]{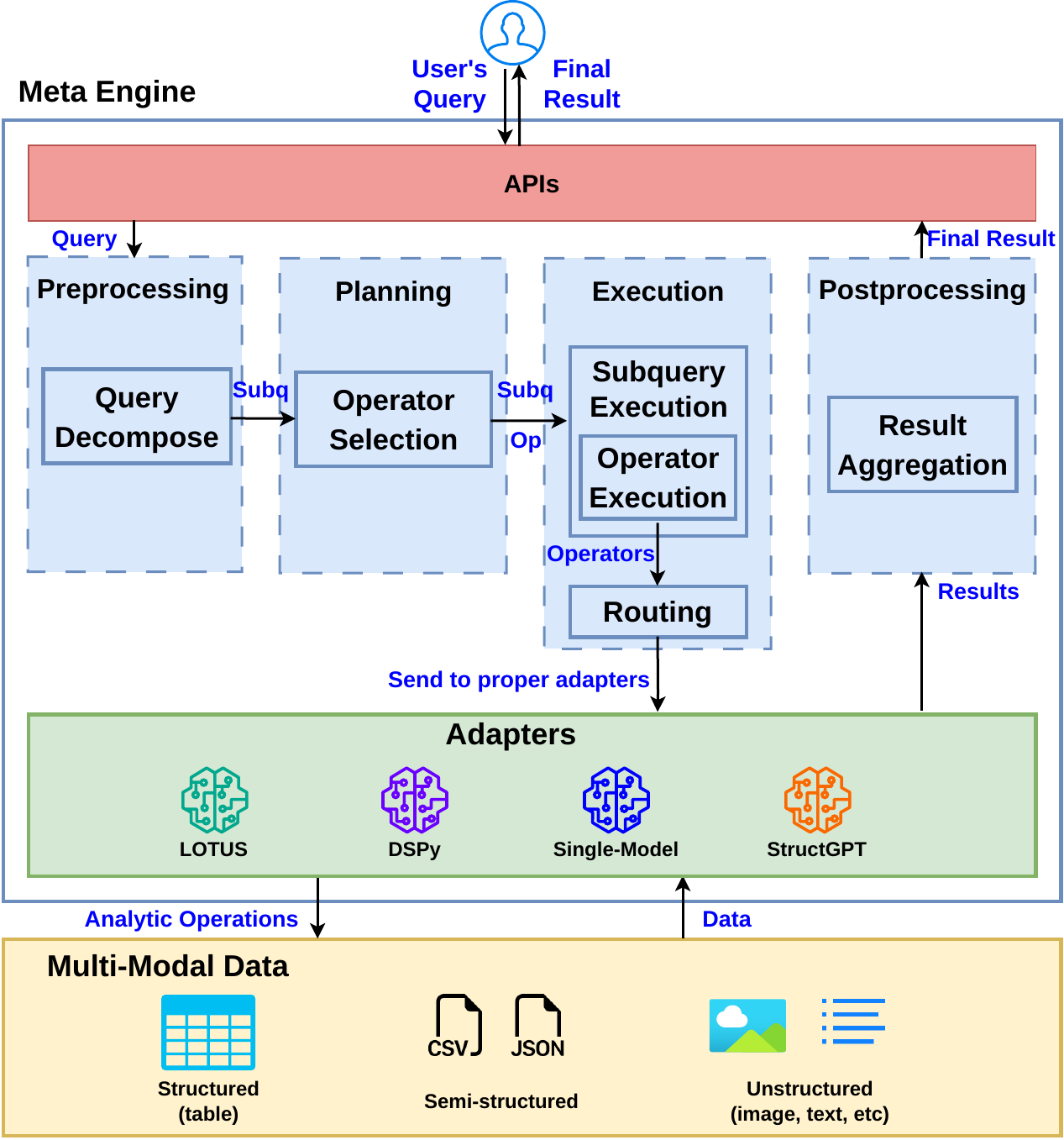}
  \vspace{-6mm}
  \caption{Architecture of Meta Engine}
  \label{fig:arch}
\end{figure}

Figure~\ref{fig:arch} presents the architecture and query/data flow of \ours{}. 

\subsection{Architecture}
\ours{} has a modular architecture that coordinates the processing of complex, multi-modal queries through specialized components. As shown in Figure~\ref{fig:arch}, the architecture includes three levels: Top-level APIs (red box), middle-level building blocks (blue boxes), and bottom-level adapters (green box). 
Users use APIs to interact with building blocks, building blocks work with each other and call adapters to process the query, while adapters access and operate the data.   

\subsubsection{APIs}
As listed in Table~\ref{tab:api}, \ours{} provides three types of APIs: (1) End-to-end query interface \texttt{query}($q: Str$), which accepts the users' natural language query and returns the final result; (2) Building block interfaces, which allows users to access the key components in \ours{} for building and customizing their pipelines, ranging from query decomposer to result refiner; (3) Data operator interfaces, which call the adapters to execute specific analytics over data in runtime. The APIs are composable and flexible that users can freely choose and combine some of them to construct a semantic query pipeline on top of multiple specialized query systems. And all APIs are extensible, i.e., \ours{} allows users to easily replace or modify the implementation of the APIs. Such APIs significantly save users' development cost, enabling efficient construction of semantic analytics pipeline on multi-modal data.

\subsubsection{Building Blocks}
\label{sec:building-blocks}
The building blocks are categorized by four stages of the query processing: query preprocessing, query planning, query execution and query postprocessing, indicated by the blue boxes in Figure~\ref{fig:arch}. The dashed boarder boxes represent the stages while the solid boarder boxes inside each stage present the components working for the stage. 

The first stage is \textbf{query preprocessing} (first blue box in Figure~\ref{fig:arch}) which checks and decomposes the input query into several subqueries. The component in this stage is the Query Decomposer. We design a dynamic decomposition algorithm that decomposes each query based on its complexity. Specifically, if the query is multi-hop, it will be decomposed to several single-hop subqueries, while if it is already single-hop, decomposition will not be conducted.  
The API for calling query decomposer is $\texttt{decompose}($q$:Str, $check$:Bool, $max\_out$:Int)$, which accepts an NL query $q$ and decomposes it into simpler sub-queries. Parameter $check$ specifies whether to enable the dynamic decomposition: if $check = True$, the query may or may not be decomposed, depending on its complexity; if $check=False$, it is forced to always decompose the query, even if it is already single-hop.
$max\_out$ specifies the maximum number of subqueries the users want. If the decomposer generates more subqueries, it will be required to re-do the decomposition with emphasis on the number of subqueries.   More details are described in Section~\ref{sec:decomposer}.

The second stage is \textbf{query planning} (second blue box in Figure~\ref{fig:arch}), which maps the NL subqueries to data operators. In our current implementation, each subquery will be mapped to one data operator among the three, i.e., \texttt{TextAnalytics}, \texttt{ImageAnalytics}, and \texttt{TableAnalytics}. Because each subquery is simple and single-hop, one data operator is enough to process it. 
The core component in this stage is Operator Generator, responsible for the Operator Selection, i.e., identifying and selecting the most appropriate analytical operators for each subquery. This mirrors the logical plan generation in DBMS, i.e., from query to logical operator(s). 
A key challenge in multi-modal data analytics is that it's unknown which modalities contain the necessary information to the query/subquery. Even though each subquery is simple, it is still hard to predict which modality of data is related to this subquery. Accessing all modalities could ensure the needed information is covered, but this solution is computation-intensive. Predicting just the most likely modality is more efficient, but wrong prediction will lead to meaningless result. To achieve maximum information coverage while maintaining good efficiency, we design a ranking-based operator selection and execution mechanism. 
Specifically, the operator generator will return a ranked list of multiple candidate data operators, rather than only selecting one operator. Then the candidate operators will be executed progressively with early stopping in execution stage. More details are introduced in Section~\ref{sec:op-selector}.
The API to call the operator generator is \texttt{select\_op}($q$: Str, $ops$: List). It accepts query $q$ and a list of available data operators (all the three data operators in \ours{} by default), returns a ranked list of candidate operators, ordered by the determination confidence. This is completed by a language model based classification.

The third stage is \textbf{query execution} (the third blue box in Figure~\ref{fig:arch}). 
As an important stage in \ours{}, query execution comprises of several components: the Query Router and various Adapters. 
An adapter is a wrapper of an existing semantic query system, such as LOTUS, DocETL, StructGPT, DSPy and so on. On one hand, these adapters encapsulate the disparate interfaces of various systems and provide unified data operator APIs for users and other components in \ours{}. This significantly saves development cost. On another hand, adapters directly interact with the multi-modal data sources (tables, text, and images) to perform the operator/subquery, ensuring the power of different systems to be jointly utilized.   

The query router sends each sub-query's data operator to the proper adapter to execute. This step mirrors the physical plan generation in DBMS, i.e., from logical operator to physical implementation, and here the physical implementation is the concrete implementation of the data operator in each underlying system. Note that each data operator is associated with the corresponding subquery (like \texttt{TextAnalytics}($q$: Str)), so the router makes the decision based on both the operator type and the subquery. Furthermore, we design two types of routers, statistics-based and learning-based, with different advantages and satisfying different demands. More details are in Section~\ref{sec:router}.    

Additionally, we propose novel optimizations for subquery execution, including iterative subquery refinement and dependency-based parallelism, which progressively refine the next subquery with previous results, and parallelize the subquery execution based on their dependencies. More details are in Section~\ref{sec:op-selector}.  

There are several APIs associated with this execution stage: \texttt{exec\_subq}($qs$: List, $par$: Bool), \texttt{exec\_op}($ops$: List, $r$: \{S$|$L\}), and  \texttt{route}($ops$: List, $adpts$: List, $r$: \{S$|$L\}). 
These APIs are implemented from top to bottom. The top level is \textit{exec\_subq} that utilizes \texttt{select\_op} and \texttt{exec\_op} to do logical planning, physical planning and execution for each given subquery. Parameter $par$ determines whether to enable the dependency-based parallelism for subqueries.
The middle level is \texttt{exec\_op}, which uses \texttt{route} to route each data operator in $ops$ to an adapter, call the adapter to execute it, and return the results.          
At the bottom, \texttt{route} predicts a proper adapter per data operator. Parameter $r$ specifies which routing algorithm to use, Statistics-based prediction ($S$) or learning-based classification ($L$).

Finally, in the \textbf{postprocessing} stage (the fourth blue box in Figure~\ref{fig:arch}), the partial results from these isolated sub-query executions are synthesized by the Result Aggregator. This component cleans the results from all sub-queries and comprehends them to produce the final result to the original user-input query.
The API to call this aggregator is \texttt{agg\_results}($res$: List, $q$: Str). It aggregates and refines the subquery results in the list $res$ into the final result to query $q$. 
An LLM will be used to conduct the aggregation. If a pipeline does not use the aggregator, the final result will be the result of the last subquery. More details are in Section~\ref{sec:agg-res}. 


With such an architecture, Meta Engine demonstrates a significant improvement in result quality compared to the state-of-the-art LLM-based query systems. By the joint use of specialized LLM systems and novel optimizations, \ours{} consistently achieves higher accuracy with competitive efficiency in our evaluation. 
Another core advantage of the architecture is its modular design. The components are flexible to be composed into customized pipelines. Each component has an API for users to directly access. The underlying LLM systems are wrapped by adapters and transparent to users. This design decouples the reasoning logic from the backend execution, allowing for seamless addition or removal of underlying systems (e.g., LOTUS, DSPy, DocETL) without overhauling the core engine, and enabling flexible pipeline construction. With new powerful LLM systems emerge, they can be easily integrated as new adapters.
We demonstrate several different pipelines built upon \ours{} in Section~\ref{sec:pipelines}.  


\subsubsection{Data Operators and Adapters}
Currently \ours{} has three data operators, shown in Table~\ref{tab:api}: (1) \textbf{TextAnalytics} operator (API \texttt{TextAnalytics}($q$: Str)) conducts analytics whose supporting evidence resides in textual data. Given a query/subquery, this operator extracts relevant facts, entities, or relations and produces the analytic results conditioned solely on the text data.
(2) \textbf{TableAnalytics} operator (API \texttt{TableAnalytics}($q$: Str)) is designed for processing analytics grounded in structured tabular data. Its processing includes filtering, row/column selection, aggregation, comparison, and numerical computation according to the query intent. 
(3) \textbf{ImageAnalytics} operator (API \texttt{ImageAnalytics}($q$: Str)) processes visual analytics that requires visual evidence from images, charts, or diagrams. It understands semantic information, including objects, labels, trends, and numeric values in the figures, and aligns these signals with the sub-query to generate results.

Each data operator has an \texttt{execute} interface in one or more adapters, which acts as the physical implementation of this operator. 
An adapter may not implement the interfaces for all data operators, depending on its speciality and limitation. 
For example, in our current implementation, StructGPT adapter only has \texttt{execute} interface for \texttt{TextAnalytics}, as StructGPT is designed to process structured data; while LOTUS, DSPy and Single-Model adapters support all three data operators.       
As a result, a data operator could be supported by multiple adapters, like LOTUS, DSPy and Single-Model adapters all support \texttt{ImageAnalytics}. In runtime the router will choose the potentially best adapter to execute the data operator, i.e., predicting and selecting the adapter which is likely to have the best performance to execute the given data operator.     

\begin{figure*}[ht]
  \centering
  \includegraphics[width=1\textwidth]{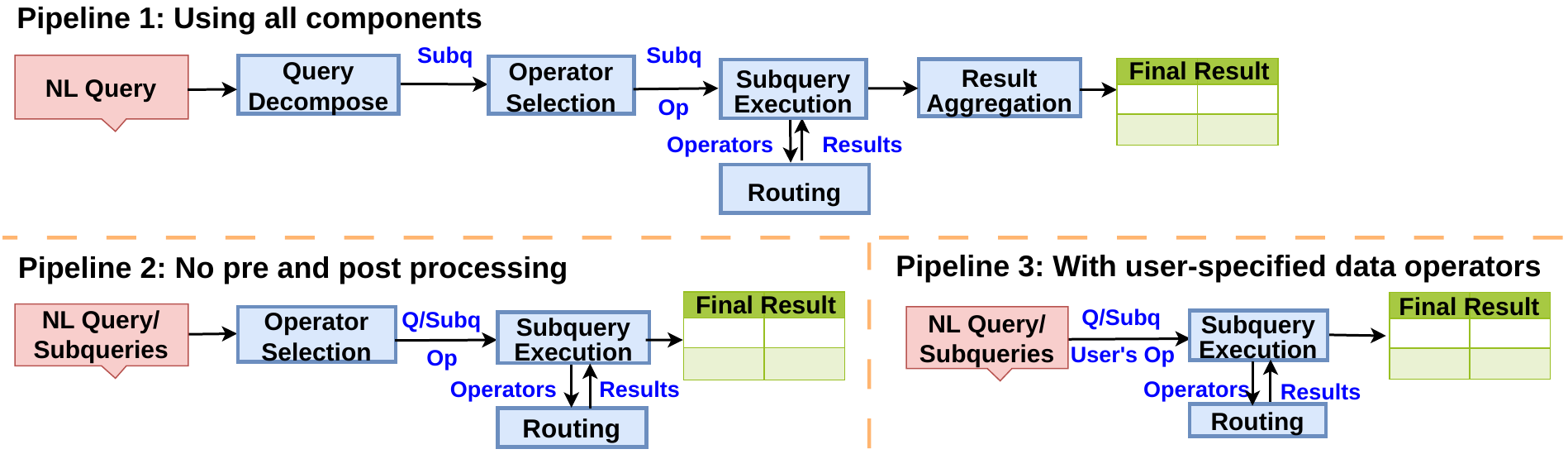}
  \caption{Example pipelines built with Meta Engine framework}
  \label{fig:pipelines}
\end{figure*}

\subsection{Pipeline Construction}
\label{sec:pipelines}
Users are free to use different components of \ours{} to customize their pipelines. Figure~\ref{fig:pipelines} demonstrates some example pipelines that can be built with \ours{} framework. To make the figures clear, we ignore the boxes of APIs and adapters. We use ``Subq'', ``Op'', ``Q/Subq'' to denote ``Subqueries'', ``Operators'' and ``Query/Subqueries''. 

Pipeline 1 is a fully automated pipeline with all the major components introduced above, from query decomposer to result aggregator. The input is the original user query, going through the whole pipeline until the final result is made. Such a pipeline provides the easiest way to use.     
Pipeline 2 is a partially automated pipeline without pre and post processing, which is for the scenarios that users have subqueries ready or prefer to keeping the query as a whole. This pipeline saves the time spent on decomposition and aggregation. 
Pipeline 3 is a mini pipeline with the minimum number of components, which gives users the most control of the query processing. Users can prepare the query/subqueries in their own way and customize the operator assignment for the query/subqueries. This pipeline only takes responsibility of executing the operators. Such a pipeline is useful in some cases, e.g., the case that users know the needed data modality. Like in a text QA task, the only needed data operator is \texttt{TextAnalytics}, thus the operator selection can be skipped to save time and tokens. In extreme cases, the router can be removed if users prefer a specific underlying system. 

In summary, \ours{} framework is highly composable and flexible. Users can build an end-to-end pipeline including all stages or customize a pipeline that starts from a specific stage.  The query preprocessing and postprocessing stages are always optional in a pipeline. Without decomposer, users can input the original query or prepared subqueries to the pipeline, while without result aggregator, the last subquery's result will be considered as the final result. Query planning stage (operator selection) can also be ignored if users prefer manually setting the data operators. The only required stage in a pipeline is the execution, including query routing and execution by adapter. Such a framework provides the maximum flexibility to pipeline construction for different demands.    
In addition, \ours{} also allows the pipelines to be controllable. Since each component is extensible, users can easily control the number and behavior of the components for pipeline diagnosis. For example, users can control the router to send all subqueries to a specific adapter by purpose, to evaluate the underlying system's performance in the pipeline.

\section{Querying Stages and Optimization}

In this section we introduce details of each components and the optimization on them.   

\begin{figure}[h]
  \vspace{-7mm}
  \centering
  \includegraphics[width=1\columnwidth]{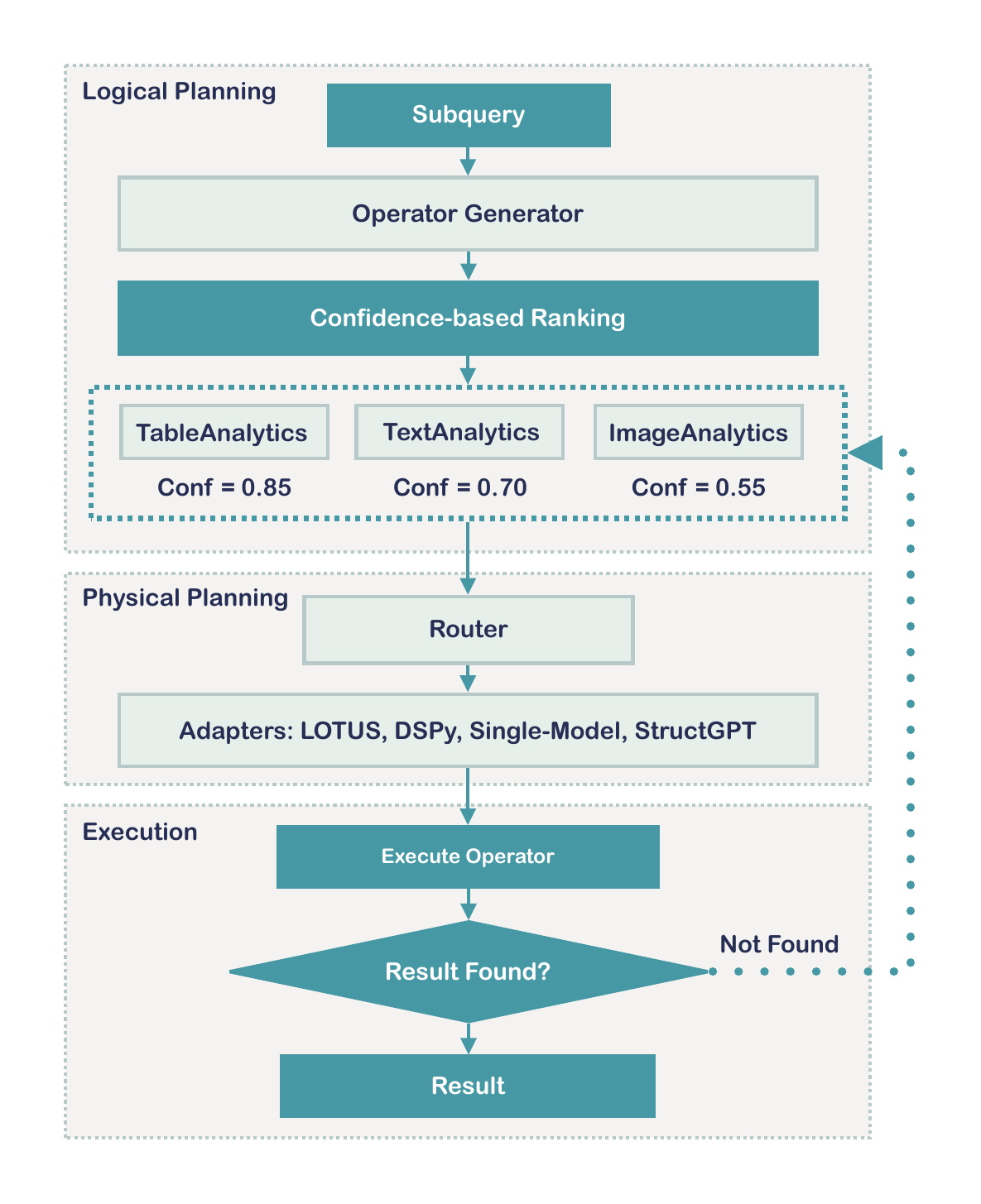}
  \vspace{-10mm}
  \caption{Confidence-based Operator Ranking, Routing, and Execution in Meta Engine}
  \setlength{\belowcaptionskip}{-50pt}
  \label{fig:op-sel}
  \vspace{-6mm}
\end{figure}

\subsection{Stage 1: Query Preprocessing}
\label{sec:decomposer}
\begin{algorithm}
\setstretch{0.1}
\small
\caption{Dynamic Query Decomposition}
\label{alg:decompose}
\begin{algorithmic}[1]
\Require{Query $q$, flag for complexity checking $F_c$, maximum subuqery number $max\_out$, maximum iterations $max\_it$} 
\Ensure{Set of subqueries $\mathcal{S}$, Dependency graph $\mathcal{G}_S$}
\State iter := 0
\State  $\mathcal{S} := \{q\}$
\While{$iter \leq max\_it$} 
    \If{$F_c = True$}
        \If{\texttt{Check}($q$) = "Complex"} 
            \State  $\mathcal{S} := \texttt{Decompose}(q)$
        \Else  
            \State  $\mathcal{S} := \{q\}$
        \EndIf    
    \Else 
        \State  $\mathcal{S} := \texttt{Decompose}(q)$
    \EndIf
    \If{$|\mathcal{S}| > max\_out$}
        \State iter := iter + 1
        \State Set feedback, emphasizing subquery number
    \Else 
        \State $\mathcal{G}_S$ := \texttt{ParseDependency($\mathcal{S}$)}
        \State \textbf{return} $\mathcal{S}$, $\mathcal{G}_S$
    \EndIf
\EndWhile
\State $\mathcal{G}_S$ := \texttt{ParseDependency($\mathcal{S}$)}
\State \textbf{return} $\mathcal{S}$, $\mathcal{G}_S$
\end{algorithmic}
\end{algorithm}
\subsubsection{Query Decomposer}

Complex queries are hard to processed accurately, therefore \ours{} provides a dynamic query decomposition method to split the input query into several simple subqueries.   
The method involves two modules, query complexity checker and the decomposer. As described in Algorithm~\ref{alg:decompose}, given an input query, the complexity checker first determines whether the query is complex enough, and for a complex query (e.g., multi-hop queries), the decomposer will split it into a set of simpler sub-queries, each corresponding to a specific task within a single data modality. 
If the checker determines the query is already simple (like single-hop), \ours{} will bypass the decomposer and treats the original query as a single sub-query. Users can turn off the checking, then all queries will be decomposed. 

Both checker and decomposer are built on top of LLM. The checker inspects whether the input query falls in any of the "Complex Categories", like multi-step reasoning, aggregation or groupby analytics, complex filtering, etc. It also looks for any "Simple" signs in the query, like whether the query includes one clear intent or asks for only one entity or fact.  
Particularly, if it is uncertain to the checker, the query is considered as complex by default.  

The resulted sub-queries follow a basic-to-advanced reasoning order, where earlier sub-queries focus on retrieval of basic/grounding facts, and later sub-queries are built upon the results of preceding ones to support higher-level inference. 
For example, in Figure~\ref{fig:exp-subquery}, the original query is decomposed into three sub-queries, where the first retrieves a basic fact about the flag while the last sub-query does advanced, higher-level analytics.    
In addition, the decomposer will ensure each sub-query can be answered within only one modality, i.e., based on one kind of data among text, images or tables. This is critical to guarantee that each subquery can be processed by one data operator.  


\subsubsection{Decomposer Optimization}
To improve the overall system efficiency, we design two optimizations: 
(1) limiting the number of sub-queries generated,  and (2) dependency-based subquery parallelism. 

Allowing unlimited decomposition may creates a significant number of sub-queries in some datasets. Those sub-queries are often unnecessarily detailed and there are often duplicates or large overlaps between them, which increases latency but does not actually produce better answers. So we allow users to specify the maximum number of subqueries they want, by the parameter $max\_out$ in Algorithm~\ref{alg:decompose}. The query decomposition will be redo if the generated subqueries exceed $max\_out$. In our evaluation, this redo rarely happens, therefore it does not slow down the decomposition.   
By limiting the number of sub-queries, we encourage the pipeline to focus on the most important analysis, reducing unnecessary computation and improving result stability for different datasets.

Another optimization is a dependency graph of subqueries, which will be used later to parallelize the subquery execution. 
Given the basic-to-advanced decomposition, later sub-queries may depend on the results of earlier ones. But the subqueries are usually not fully dependent. 
For example, a query asking ``Which person has a lower salary in 2025: the person in blue with a dog in the picture XXX, or the person with the highest-level position in company YYY'' could be decomposed into three subqueries: (1) "What is the salary of the person in blue with a dog in the picture XXX", (2) "What is the salary of the person with the highest-level position in company YYY", and (3) "Which person has a lower salary between them". In this example, the subquery 3 relies on subqueries 1 and 2, while subqueries 1 and 2 have no dependency between each other.      
Such partially independent subqueries can be parallely executed to improve the pipeline efficiency.   
To handle such dependencies, the decomposer further organizes the subqueries into a dependency graph whose nodes are subqueries and edges are dependency between subqueries, which is used in execution stage.














\begin{figure}[t]
\centering
\small
\setlength{\fboxsep}{6pt}
\fbox{\parbox{0.93\columnwidth}{
\textbf{Original Query:} \textit{``When was the last time capital punishment took
place in the state whose flag features a bear and
has a major city in the Sun Belt?''}

\vspace{4pt}
\hrule
\vspace{4pt}

\textbf{Sub-query 1} {\scriptsize [image]}\\
\textit{``Which state has a flag that features a bear?''}\\
$\rightarrow$ \texttt{imageAnalytics("Which state has a flag that
features a bear?", ['California'])}\\
\textbf{Result:} California

\vspace{3pt}

\textbf{Sub-query 2} {\scriptsize [table]} {\color{gray}\scriptsize $\leftarrow$ depends on Sub-query 1}\\
\textit{``Does this state have a major city located
in the Sun Belt region?''}\\
$\rightarrow$ \texttt{tableAnalytics("Does California have a
major city located in the Sun Belt region?",
['Sun Belt'])}\\
\textbf{Result:} Anaheim, Bakersfield, Fresno, Long Beach, Los Angeles, Oakland, Riverside, Sacramento, San Bernardino, San Diego, San Jose, San Francisco

\vspace{3pt}

\textbf{Sub-query 3} {\scriptsize [text]} {\color{gray}\scriptsize $\leftarrow$ depends on Sub-query 1 and 2}\\
\textit{``When was the last time capital
punishment took place in this state?''}\\
$\rightarrow$ \texttt{textAnalytics("When was the last time capital
punishment took place in California?",
['Capital punishment in California'])}\\
\textbf{Result:} 2006 $\Rightarrow$ \textbf{Final Answer:} 2006 in California.
}}
\caption{An example of subquery decomposition and dependency parsing, operator selection, execution. }


\label{fig:exp-subquery}
\vspace{-10pt}
\end{figure}

\subsection{Stage 2 \& 3: Query Planning and Execution}
\label{sec:op-selector}

After sub-queries are produced, the next stages are logical planning, physical planning and execution, which is shown in Figure~\ref{fig:op-sel}. 

In \ours{}, the logical planning is conducted by a confidence-based operator ranking. As mentioned in Section~\ref{sec:building-blocks}, it is hard to accurately predict which data modality includes the necessary information to process a subquery. So instead of predicting the most proper data operator, \ours{} ranks the available data operators (TextAnalytics, ImageAnalytics, and TableAnalytics in this paper) by prediction confidence, and execute them progressively later.     
Instead of using full data contents, this ranking relies only on metadata like document titles and short captions (e.g., one-sentence summaries for images), to keep the planning efficient and low-cost. 
Given the metadata and subquery, each candidate operator is assigned a confidence score (from 0 to 1) by the LLM-based Operator Generator, which estimates how suitable the modality and operator is for processing the subquery. The ranking will consider whether the subquery has direct mention or implication of the modality, whether the requested information refers to any specific modality, and so on. The scored candidate operators are sorted in descending order of confidence.

Then in physical planning, each data operator will be mapped to a specific adapter's interface by the query router.  
The query router intelligently maps each operator to the most appropriate underlying specialist system. Note that each operator could have multiple choices, like LOTUS, DSPy and VLM can all execute \texttt{ImageAnalytics}. In these cases the router will predict the best-performing adapter on the given operator and subquery. 
For this goal, we design two types of routers: a learned router which is a query-dependent classifier on top of a pre-trained LLM, and a statistic router that is query-agnostic, purely based on history statistics. Our evaluation shows the learned router is overall more accurate, but statistic router outperforms it in some cases. We provide more details in Section~\ref{sec:router}. 

After the physical planning, each candidate operator in the ranked list has an adapter to execute it. There are two levels of execution, operator execution and subquery execution. At operator level, the ranked operators will be executed by their adapters progressively: they are executed in order (higher confidence operator first). If an operator does not return valid/meaningful result, next operator will be executed, until meaningful result is returned. This progressive execution ensures the processing quality, since the top-ranked operator may not always be the most proper one for the subquery. Figure~\ref{fig:exp-subquery} shows an example on the operator that returns valid results for each subquery. 
In our evaluation, the average number of executed operators per subquery ranges from 1.05 to 1.2, proving that this progressive execution raises only negligible overhead.

Executing a subquery is essentially executing its data operators as above. Additionally, at subquery level, the execution is optimized with two techniques: iterative subquery refinement and dependency-based parallelization. The dependency graph generated in decomposition is used by both techniques.  
First, for sub-queries after the first one, \ours{} additionally refines them before execution. Specifically, next sub-query is refined by incorporating execution results from previous sub-queries, such as filling the resolved entity names or intermediate values from the results into next sub-query to resolve abstract references into concrete entities and reduce ambiguity.
Before executing the $i$-th sub-query, the results from sub-queries $1$ through $i{-}1$ which the $i$-th sub-query has dependency to are provided as additional context. With such additional context, the $i$-th sub-query will be refined as above by LLM. Figure~\ref{fig:exp-subquery} shows an example on how the subqueries are refined, e.g., the result ``California'' from subquery 1 is filled into subquery 2, shown in its \texttt{TableAnalytics} operator.
By such a dynamic refinement, the sub-queries are stated more clearly and the system avoids wasting time in duplicate computation across sub-queries

Second, execution of subqueries is parallelized based on their dependencies. Given the dependency graph, \ours{} first conduct a BFS-like search to split the graph into stages based on the depth, where each subquery in a stage is dependent to some subqueries in previous stages, and all subqueries in the same stage are independent from each other. Based on such split, the subqueries are executed sequentially across stages while parallelly within each stage. Such a dependency-based parallelization ensures a competitive execution efficiency given the relatively slow LLM calls.


\subsection{Stage 3: Aggregator and Final Result}
\label{sec:agg-res}

\label{subsec:adapters}

This final component aggregates the individual pieces of information and generates the final results. It refines the results from different subqueries (like deduplicating the overlaps between results) and synthesize them to the comprehensive and accurate final result. 
Ideally, the result of the last subquery is the final result to the original query. But there could be information conflicts between the results of last subquery and previous subqueries. So a major responsibility of the aggregator is to resolve conflicts between them by checking logical consistency across the reasoning chain and preferring the result that are coherent with previously established intermediate facts. 
This aggregator is optional in a pipeline. Without it, the final result will be the result from the last subquery by default.

\subsection{Query Router}

\label{sec:router}
In this and the next section we introduce two components, the router and adapters. 
For each operator, we utilized a router to select the most appropriate adapter to execute it. 
We construct two different types of routers, statistic and learned routers.

\textbf{Statistic router} is based on the statistics of best adapter for each operator over an annotated training set. Specifically, for the operator used in a training sub-query, we try all capable adapters (one by one) to execute the operator,  and collect the performance metrics (e.g., \textit{Semantic Hit}, introduced in Section~\ref{sec:exp-settings}) of each adapter's execution. Then for real (testing) queries, each operator will be routed to the adapter with the best overall performance for it across all training sub-queries. In short, the statistic router will always route an operator to the best-performing adapter for it based on history results. We denote this statistic, static, sub-query agnostic router as \textit{StatRouter}.

\textbf{Learned router} (denoted as \textit{MLRouter}) is a learning-based, dynamic, sub-query dependent router.  
It accepts the sub-query, operator, and modality indicators as input, and learns to perform semantic-aware routing based on the sub-query and operator. 
We implement the learned router using Qwen-0.6B\cite{qwen3technicalreport} as a frozen encoder, followed by a lightweight multilayer perceptron (MLP) classifier. Only the parameters of the MLP are trained, while the backbone language model (Qwen) remains fully frozen throughout training and inference. Router training is conducted offline using historical execution traces collected from all the adapters over the training sub-queries. 
Specifically, for each sub-query and its operator, all capable adapters are executed individually, and their outputs are evaluated using multiple metrics, including \textit{F1}, \textit{Hit}, \textit{Coverage}, and \textit{Semantic Hit} (all introduced in Section~\ref{sec:exp-settings}). 
These metric scores are combined into a single quality score by weighted summation, where Semantic Hit and F1 scores are emphasized, while Hit and Coverage serve as secondary criteria. 
For each training sub-query $q$, the training label is the best adapter $k^*_q$ for executing this sub-query. 
With such training data, the router can learn to choose the proper adapter dynamically based on the sub-query.        
The training is performed using a weighted cross-entropy objective, where each training sub-query contributes proportionally to its sample weight. The sample weight for a training sub-query is computed as a function of (1) the absolute quality score of the best-performing adapter and (2) the margin between the best and sub-optimal adapters, as shown in Equation~\ref{eq:sample-weight} and Equation~\ref{eq:weighted-ce} . 

Formally, let $S_q^k \in [0,1]$ denote the aggregated quality score of adapter $k$
when executing sub-query $q$.
Let $k^*_q = \arg\max_k S_q^k$ be the best-performing adapter, and define
$S_q^{(1)} = \max_k S_q^k$ and $S_q^{(2)}$ as the second-best score among all adapters.
Each training sub-query $q$ is associated with a confidence-aware sample weight
$w_q$, defined as
\begin{equation}
\label{eq:sample-weight}
w_q
=
\min\!\Big(
w_{\max},
\max\!\big(
w_{\min},
\alpha + \beta S_q^{(1)} + \gamma (S_q^{(1)} - S_q^{(2)})
\big)
\Big).
\end{equation}
The weight is bounded within $[w_{\min}, w_{\max}]$ to prevent overly dominant or
vanishing contributions from individual training records.
In our experiments, we set $\alpha=0.1$, $\beta=0.9$, $\gamma=0.2$,
$w_{\min}=0.1$, and $w_{\max}=1.5$.

Given the router prediction $p_\theta(k \mid q)$ over adapters,
training is performed using a weighted cross-entropy objective, to maximizes $p_\theta(k^*_q \mid q)$, i.e., the predicted probability of the groundtruth best adapter $k^*_q$ given sub-query $q$:
\begin{equation}
\label{eq:weighted-ce}
\mathcal{L}
=
\frac{1}{\sum_q w_q}
\sum_q
w_q \cdot
\left(
-\log p_\theta(k^*_q \mid q)
\right).
\end{equation}

Such an objective considers not only the best adapter but also the sub-optimal adapters whose performance is close to the best one, which makes the router more robust, as the best adapter in training may not be still the best in testing, but it is likely still among the top adapters. 
The weighting mechanism allows the router to emphasize reliable routing signals without requiring hard filtering or manual thresholding of training data\cite{shazeer2017outrageously}. 

During inference, the router performs a single forward pass to produce a categorical adapter decision for each sub-query. As a result, both the statistic and learned routers raise negligible overhead of latency comparing to the downstream sub-query execution.

The routing module is fully decoupled from adapter implementations and downstream reasoning logic. New adapters can be added or removed by extending/shrinking the classification label space and retraining only the lightweight classifier, without modifying the backbone model or the rest of the system, which makes \ours{} practical and adaptive to dynamic user demands.

\subsection{Adapters}
\label{sec:adapters}

The adapters wrap the underlying semantic query systems (e.g., LOTUS, DSPy, etc) and provide unified APIs for executing the operators of \ours{}.
The unified interfaces allows transparently calling the diverse underlying systems.

Most of the adapters support all the three analytical operators in \ours{}, i.e., TextAnalytics, TableAnalytics and ImageAnalytics, except StructGPT adapter which only supports TableAnalytics.  
Specifically, each operator has one interface (entry function) in each adapter which supports it. 
The interface accepts a sub-query and returns the execution result of the current operator over the sub-query.

Currently \ours{} integrates four adapters, warping LOTUS, DSPy, Single-Model and StructGPT respectively.  In this paper, we enforce deterministic decoding (temperature $=0$, fixed seed) to improve reproducibility and system stability.

\textbf{LOTUS adapter} supports all three modality operators. We implement each interface mainly using the semantic aggregation (\textit{sem\_agg}) in LOTUS. The supporting documents are stored in Pandas Dataframe by LOTUS, and \textit{sem\_agg} is called over them to comprehend the evidence of corresponding modality and analyze the sub-queries. 

\textbf{DSPy adapter} implements each modality operator with DSPy, a programmatic prompting framework that treats each operator call as an explicit \emph{Language Model (LM) Program}.  DSPy factors the operator into reusable components (e.g., instruction template and constrained output format) and can compile these components into a stable program for repeated invocation across queries. 

\textbf{Single-model adapter} implements the three modality operators via direct end-to-end prompting of a single vision-language model (VLM), serving as a strong baseline. For each operator, the adapter packages the supporting evidence in the corresponding modality and queries the VLM in one pass. 
In contrast to the other backend systems, the single-model adapter relies purely on the VLM’s ability to aggregate evidence and generate the final answer.

\textbf{StructGPT adapter} implements TableAnalytics operator with a structure-aware, multi-stage inference procedure that explicitly exploits the schema and row organization of the tables. 
For each sub-query, the adapter first performs \emph{schema grounding} to select relevant columns for the sub-query, then it applies \emph{row filtering} over the reduced schema, identifying candidate records that are likely to contain the answer. Finally, the adapter produces a response based on the candidate records. 

The adapter framework is extensible: new backend systems can be easily integrated by wrapping into new adapters, providing the required interfaces and following the standard output format, without modifying the core execution pipeline. The router can also adapt to the new adapters with minimum cost, as discussed in Section~\ref{sec:router}.


\begin{table}[t]
\centering
\caption{Statistics of the datasets}
\vspace{-10pt}
\label{tab:dataset}
\resizebox{\columnwidth}{!}{%
\begin{tabular}{lccccc}
\toprule
\textbf{Dataset} & \textbf{Queries} & \textbf{Text Docs} & \textbf{Tables} & \textbf{Images} & \textbf{Scenario} \\
\midrule
MultiModalQA & 23,817 & 218,285 & 10,042 & 57,058 & Multi-hop Q\&A \\
ManyModalQA  & 1,440  & 1,440   & 1,440  & 1,440  & Single-hop Q\&A \\
Text2Vis     & 1,788  & 1,788   & 1,788  & 1,788  & Analytical \\
FinMMR       & 1,291  & 1,291   & 0      & 2,269  & Analytical \\
M2QA         & 1,000  & 1,884   & 4,686  & 37,972 & Multi-hop Q\&A \\
\bottomrule
\end{tabular}%
}
\vspace{-10pt}
\end{table}

\begin{table*}[t]
\centering
\caption{Main experiments results across 5 datasets. Metrics are reported as percentages (\%).}
\vspace{-10pt}
\label{tab:all_results}
\resizebox{\textwidth}{!}{
\begin{tabular}{l|l|cccccc}
\toprule
Dataset & Metric 
& Lotus 
& Single-Model 
& LlamaIndex 
& DSPy 
& Meta-Engine (MLRouter) 
& Meta-Engine (StatRouter) \\
\midrule

\multirow{4}{*}{Text2Vis}
& F1-score      & 9.71  & 32.18 & 9.62  & 23.45 & \underline{\textbf{36.76}} & 36.14 \\
& Hit           & 58.11 & 64.32 & \underline{\textbf{69.74}} & 63.03 & 66.28 & 65.83 \\
& Semantic Hit  & 63.48 & 75.11 & 73.66 & 68.79 & 74.78 & \underline{\textbf{75.62}} \\
\midrule

\multirow{4}{*}{ManyModalQA}
& F1-score      & 9.48  & 33.90 & 12.89 & 33.38 & 51.06 & \underline{\textbf{54.61}} \\
& Hit           & 68.40 & 69.93 & 69.24 & 68.13 & 67.22 & \underline{\textbf{74.38}} \\
& Semantic Hit  & 68.82 & 77.78 & 74.72 & 72.78 & 72.22 & \underline{\textbf{81.60}} \\
\midrule

\multirow{4}{*}{MultiModalQA}
& F1-score      & 9.47  & 34.88 & 10.03 & 24.26 & \underline{\textbf{54.71}} & 53.13 \\
& Hit           & 70.10 & 70.95 & 65.50 & 64.15 & \underline{\textbf{75.15}} & 74.90 \\
& Semantic Hit  & 53.90 & 65.55 & 57.60 & 58.90 & \underline{\textbf{72.90}} & 72.65 \\
\midrule

\multirow{4}{*}{M2QA}
& F1-score      & 5.43  & 26.67 & 10.14 & 28.27 & 47.91 & \underline{\textbf{50.57}} \\
& Hit           & 46.11 & 43.51 & \underline{\textbf{77.25}} & 77.05 & 65.07 & 69.86 \\
& Semantic Hit  & 38.72 & 39.12 & 62.08 & 64.87 & 62.87 & \underline{\textbf{66.47}} \\
\midrule

\multirow{4}{*}{FinMMR}
& F1-score      & 1.16  & 20.75 & 2.92  & 19.70 & 26.90 & \underline{\textbf{28.47}} \\
& Hit           & 11.62 & 25.41 & \underline{\textbf{42.37}} & 25.02 & 35.40 & 37.41 \\
& Semantic Hit  & 16.42 & 34.86 & \underline{\textbf{53.99}} & 34.24 & 50.43 & 52.59 \\
\bottomrule
\end{tabular}
}
\vspace{-5pt}
\end{table*}

\section{Experiments}
\label{sec:exps}

\subsection{Experiment settings}
\label{sec:exp-settings}

All experiments are evaluated on a machine with one NVIDIA B200 80GB Tensor Core GPU, 3 AMD Rome cores CPU and 20GB RAM.


\subsubsection{Datasets}
\label{subsec:datasets}

We evaluate \ours{} on several publicly available multi-modal datasets: MultiModalQA~\cite{talmor2021multimodalqa}, ManyModalQA~\cite{hannan2020manymodalqamodalitydisambiguationqa}, Text2Vis~\cite{rahman2025text2vischallengingdiversebenchmark}, M2QA ~\cite{abaskohi2025fmds} and FinMMR~\cite{tang2025finmmrmakefinancialnumerical}.
Table~\ref{tab:dataset} summarizes their statistics.
All datasets involve supporting data from at least two of the three modalities: text, tables, and images. Each query in those datasets is supposed to be answered based on the supporting textual, tabular, and/or visual information. And we unify the supporting documents from all datasets into a standardized multimodal document repository. 
 
\textbf{ManyModalQA} is a question-answering (QA) dataset including questions, supporting data from the three modalities above, and ground-truth answers, where many questions are single-hop. 

\textbf{MultiModalQA} is another QA dataset including questions, answers and multimodal supporting data, too. But its questions are usually multi-hop.   

\textbf{Text2Vis} is a multi-modal analytics dataset, where each record includes a natural language query for a analytic task, as well as supporting documents: data tables, background introduction about the data, and charts based on the tables.  
Particularly, the charts are generated by ourselves. The original Text2Vis dataset includes Python code for visualizing the key calculation/statistics based on the table and query. We run the code to generate the charts as  supporting images.  

\textbf{FinMMR} is a multimodal benchmark focusing on financial numerical reasoning, where each query requires integrating financial knowledge with visual evidence such as tables, charts, and structured diagrams. All questions demand precise numerical results. 

\textbf{M2QA} is a multimodal question answering dataset constructed from real-world Wikipedia articles, where each record is associated with a question and a source Wikipedia URL, where the question needs to be answered based on the Wiki page. To use this dataset, we retrieve and separate different modality data from the referenced Wiki pages. The extracted text, images, and tables are then normalized into our unified multimodal document repository. In the evaluation on this dataset, we skip the queries whose supporting images are not fully valid, caused by retrieval issues in preprocessing. Finally there are 501 queries used in the evaluation.

Note that in all experiments, due to the non-trivial latency of LLM-based systems, we use subsets of each dataset rather than their full collections. For MultiModalQA we randomly sample 2000 queries from the 23817 queries, while for ManyModalQA, Text2Vis, and FinMMR, we use the whole datasets. For M2QA, as mentioned above, we use 501 queries.





\subsubsection{Methods}
We evaluate our approach against several representative baselines: Single-Model, DSPy~\cite{dspy-khattab2023dspycompilingdeclarativelanguage}, LOTUS~\cite{lotus-patel2024lotusenablingsemanticqueries}, and LlamaIndex~\cite{Liu_LlamaIndex_2022}. All baselines follow a unified implementation pattern for multiple modalities. Specifically, since the baseline systems like LOTUS and DSPy usually do not well handle cross-modal interactions (like processing text and image together for one query), we implement the baseline methods by separating-then-aggregating: Each input query is processed within each single modality of the three, i.e., analyzing it with supporting text, images and tables separately. Then the results from the three modalities will be comprehended by an LLM into the final result. All methods, including \ours{}, are using GPT-4.1-mini as the backbone LLM in this paper. 
Implementations for each baseline are shown below:    

\textbf{Single Model} baseline is a single multi-modal LLM, i.e., GPT-4.1-mini in this paper.  

\textbf{LOTUS} is a Pandas~\cite{reback2020pandas} DataFrame based semantic query system, with semantic operators powered by LLM. In this baseline, we mainly use LOTUS's semantic aggregation operator ($sem\_agg$) for processing each modality and final aggregation. This baseline will aggregate the supporting documents in each modality, get the results, and aggregate the results into the final result. 

\textbf{DSPY} is a programmatic prompting framework that lets users define AI pipelines as modular programs (signatures + modules) and execute them with an underlying LLM/VLM backend. In this baseline, 
we instantiate a \textit{ChainOfThought} module to analyze the original question separately for each modality by conditioning only on the corresponding supporting evidence. 
Then, we use a lightweight \textit{Predict} module as an aggregator to generate the final response. 

\textbf{LlamaIndex} is a programming framework for building LLM-based workflows over heterogeneous data sources. In this baseline, we use its indexing and querying abstractions to implement a retrieval-augmented generation (RAG) workflow for text and table modalities, and a direct vision-language querying workflow for images. 
The final answer is generated by synthesizing the modality-level outputs with an LLM. 

\textbf{Meta-Engine}, as we mentioned in the previous parts, 
we build two pipelines with \ours{} framework: \textbf{Meta-Engine (MLRouter)} and \textbf{Meta-Engine (StatRouter)}. Both use all the components and identical configurations (as the Pipeline 1 in Figure~\ref{fig:pipelines}), differing only in the routing strategy. MLRouter is the learned router, while StatRouter is the statistic router. 

\subsubsection{Metrics}
We evaluate end-to-end answer quality using token-based and semantic similarity metrics.
For datasets with multiple acceptable ground-truth answers per question,
we compute each metric against every ground-truth answer and report the maximum score.

Let $\hat{y}$ denote the predicted answer and $y$ a ground-truth answer.
After normalization (lowercasing, punctuation removal, whitespace trimming),
we tokenize both into token multisets $T(\hat{y})$ and $T(y)$,
and write $m(\hat{y},y)=|T(\hat{y}) \cap T(y)|$ for the multiset overlap count.
We compute the standard \textbf{token-level F1} score from the token-overlap precision and recall.
We also use \textbf{Hit and Semantic Hit}.
\emph{Hit} is a binary indicator of whether $\hat{y}$ shares at least one token with $y$, i.e., $\mathrm{Hit}(\hat{y},y)=\mathbb{I}[m(\hat{y},y)>0]$.
However, token-overlap metrics may penalize valid paraphrases or miss subtle factual errors.
We therefore additionally employ an LLM as a semantic judge:
given the question, $\hat{y}$, and the ground-truth answer set,
the judge determines whether $\hat{y}$ conveys the same factual content as at least one ground-truth answer,
ignoring surface-form variations.
We define $\mathrm{Sem\_Hit}=1$ if the judge deems $\hat{y}$ semantically equivalent, and $0$ otherwise.
We report average Hit and average Semantic Hit (i.e., fraction of questions achieving a hit) across all questions.
We also measure end-to-end query processing time as an efficiency metric.

\subsection{End-to-End Evaluation}
\label{sec:end2end-exp}

Across all datasets, Meta Engine achieves the strongest end-to-end result quality in most cases, showing improvements over all evaluated baselines across a wide range of scenarios, particularly on queries requiring multi-step reasoning and cross-modal evidence integration like MultiModalQA. 

In terms of both F1-score and semantic correctness, Meta Engine exhibits clear and substantial advantages. These results indicated that advantages are not confined to a single dataset or modality, but generally across all datasets. In a few cases, LlamaIndex achieves higher Hit scores, but its Semantic Hit and F1-score are consistently lower than that of Meta Engine. This suggests that LlamaIndex tends to produce long, verbose answers which are incorrect semantically, making no sense, while Meta Engine delivers more semantically accurate results.

Specifically, according to Table~\ref{tab:all_results}, Meta Engine shows clear gains over the strongest competing baselines on most datasets.
On Text2Vis, Meta Engine (MLRouter) achieves an F1-score of 36.76, exceeding the best baseline (Single-Model, 32.18) by 4.6\%.
On ManyModalQA, Meta Engine (StatRouter) reaches an F1-score of 54.61, outperforming the strongest baseline by 20.7\%, while improving Semantic Hit from 77.78 to 81.60.
On MultiModalQA, Meta Engine (MLRouter) attains an F1-score of 54.71, surpassing the best baseline by 19.8\%, together with a Semantic Hit increase of over 7\%.
On M2QA, Meta Engine (StatRouter) achieves an F1-score of 50.57, improving upon the strongest baseline by 22.3\%, and yields the highest Semantic Hit (66.47).
On FinMMR, Meta Engine attains an F1-score of 28.47, which is comparable to LlamaIndex, while remaining competitive in terms of semantic correctness across all baselines.

Existing baselines typically attempt to answer complex queries in a single pass or through a fixed pipeline, forcing a single model or framework to handle heterogeneous reasoning requirements simultaneously. In contrast, Meta Engine decomposes complex queries into simpler, modality-aligned sub-queries, ensuring that each reasoning step is grounded in the most appropriate data modality and reducing cross-modal interference during reasoning. Also, Meta Engine overcomes this limitation by dynamically routing each sub-query to the most appropriate adapter, effectively combining the strengths of multiple specialized systems. This coordination mechanism enables Meta Engine to leverage modality-specific expertise during execution, leading to consistently stronger overall performance.

While some baselines achieve moderately higher hit rates through more token overlap with ground-truth answers, they often produce incomplete, verbose, or weakly grounded responses. In contrast, Meta Engine consistently attains higher semantic accuracy, indicating that its answers better capture the intended factual content of the queries. These improvements come from the effective \ours{} architecture of query decomposition, operator generation, routing, and multi-adapter execution.   

Furthermore, we apply a lightweight post-processing step for the baselines which tends to output verbose answers (such as LOTUS), to removing the leading explanatory text and extracts the major content from their answers. However, even after this step, the resulting baseline answers still remain long, reflected on the low F1 scores of LOTUS and LlamaIndex. This proves that simple addition to the existing semantic query systems cannot effectively improve them, which strengthens the necessity of the new architecture and workflow in \ours{}.  

\begin{table}[t]
\centering
\small
\caption{Ablation results for MultiModalQA. All metrics are reported as percentages (\%).}
\vspace{-10pt}
\label{tab:multimodalqa_ablations}
\begin{tabular}{l|ccc|H}
\toprule
\multicolumn{5}{c}{\textbf{MultiModalQA (len = 2000)}} \\
\midrule
Method & F1 (\%) & Hit (\%) & Semantic Hit (\%) & Runtime (s) \\
\midrule

\multicolumn{5}{l}{\textbf{MLRouter}} \\
\midrule
Meta-Engine               
& 54.71 & 75.15 & 72.90 & 11.91 \\

Meta-Engine w/o Aggregator 
& 52.75 & 74.15 & 70.60 & 11.76 \\

Meta-Engine w/o Decomposer
& 54.66 & 72.70 & 71.40 & 7.33 \\


Meta-Engine w/o Checker
& 52.60 & 76.10 & 71.95 & 19.26 \\

\midrule
\multicolumn{5}{l}{\textbf{StatRouter}} \\
\midrule
Meta-Engine     
& 53.13 & 74.90 & 72.65 & 13.08 \\

Meta-Engine w/o Aggregator 
& 40.79 & 75.15 & 71.60 & 12.26 \\

Meta-Engine w/o Decomposer
& 53.01 & 73.05 & 71.00 & 9.02 \\


Meta-Engine w/o Checker
& 53.18 & 76.80 & 73.85 & 18.74 \\

\midrule
\multicolumn{5}{l}{\textbf{Single-Adapter (LOTUS)}} \\
\midrule
Meta-Engine               
& 39.76 & 62.35 & 57.15 & 13.78 \\
\bottomrule
\end{tabular}
\end{table}













\begin{table}[t]
\centering
\small
\caption{Ablation results for ManyModalQA. All metrics are reported as percentages (\%).}
\vspace{-8pt}
\label{tab:manymodalqa_ablation}
\begin{tabular}{l|ccc|H}
\toprule
\multicolumn{5}{c}{\textbf{ManyModalQA (len = 1440)}} \\
\midrule
Method & F1 (\%) & Hit (\%) & Semantic Hit (\%) & Runtime (s) \\
\midrule
\multicolumn{5}{l}{\textbf{MLRouter}} \\
\midrule
Meta-Engine            
& 51.06 & 67.22 & 72.22 & 8.15 \\

Meta-Engine w/o Aggregator 
& 50.50 & 68.19 & 72.50 & 6.45 \\

Meta-Engine w/o Decomposer
& 51.47 & 68.19 & 72.64 & 5.76 \\


Meta-Engine w/o Checker
& 51.09 & 72.78 & 76.67 & 13.04 \\

\midrule
\multicolumn{5}{l}{\textbf{StatRouter}} \\
\midrule
Meta-Engine              
& 54.61 & 74.38 & 81.60 & 8.72 \\

Meta-Engine w/o Aggregator 
& 46.80 & 74.86 & 81.25 & 7.53 \\

Meta-Engine w/o Decomposer
& 54.40 & 74.38 & 80.14 & 6.12 \\


Meta-Engine w/o Checker
& 51.00 & 73.47 & 78.89 & 14.53 \\

\midrule
\multicolumn{5}{l}{\textbf{Single-Adapter (LOTUS)}} \\
\midrule
Meta-Engine
& 34.83 & 53.54 & 57.85 & 9.26 \\

\bottomrule
\end{tabular}
\vspace{-5pt}
\end{table}

\begin{table}[t]
\centering
\small
\caption{Ablation results for M2QA. All metrics are reported as percentages (\%).}
\vspace{-10pt}
\label{tab:m2qa_ablation}
\begin{tabular}{l|ccc|H}
\toprule
\multicolumn{5}{c}{\textbf{M2QA (len = 501)}} \\
\midrule
Method & F1 (\%) & Hit (\%) & Semantic Hit (\%) & Runtime (s) \\
\midrule
\multicolumn{5}{l}{\textbf{MLRouter}} \\
\midrule
Meta-Engine             
& 47.91 & 65.07 & 62.87 & 19.89 \\

Meta-Engine w/o Aggregator 
& 35.62 & 59.08 & 52.50 & 17.00 \\

Meta-Engine w/o Decomposer 
& 47.47 & 60.28 & 58.48 & 8.30 \\


Meta-Engine w/o Checker
& 43.88 & 65.47 & 61.28 & 23.58 \\

\midrule
\multicolumn{5}{l}{\textbf{StatRouter}} \\
\midrule
Meta-Engine           
& 50.57 & 69.86 & 66.47 & 21.01 \\

Meta-Engine w/o Aggregator 
& 29.20 & 66.27 & 57.29 & 20.99 \\

Meta-Engine w/o Decomposer 
& 51.02 & 68.86 & 65.47 & 10.46 \\


Meta-Engine w/o Checker
& 45.80 & 67.86 & 64.67 & 25.81 \\

\midrule
\multicolumn{5}{l}{\textbf{Single-Adapter (LOTUS)}} \\
\midrule
Meta-Engine 
& 46.58 & 69.06 & 65.27 & 20.93 \\
\bottomrule
\end{tabular}
\vspace{-10pt}
\end{table}

\subsection{Ablation Study}
\label{sec:ablation-exp}
To understand the contribution of individual components in the Meta-Engine, we conduct a series of ablation experiments by selectively disabling specific modules. Each ablation isolates a design choice and reveals its functional role in the overall system. The results are shown in Table~\ref{tab:multimodalqa_ablations}, \ref{tab:manymodalqa_ablation} and \ref{tab:m2qa_ablation}. 

\textbf{Without Checker: }
In this setting, the complexity checker is disabled, and all queries are treated as complex by default. As a result, every input query is decomposed into sub-queries, regardless of whether decomposition is actually necessary. 
Without checker, the result quality is usually lower in most cases. Although there are a few exceptions (Table~\ref{tab:multimodalqa_ablations}), this does not affect the necessity of the checker, as checker improves the end-to-end efficiency significantly. As in Table~\ref{tab:runtime_all_ablations}, decomposing all queries without the checker could raise up to 70\% more processing time. By preventing decomposing simple queries, a non-trivial portion of queries are processed in only one pass.      

\textbf{Without Decomposer: }
In the w/o Decomposer setting, the system never performs query decomposition. All original questions are considered non-complex and are passed directly into the pipeline as a single sub-query.
Removing decomposer makes the system overall perform worse, proving the effectiveness of query decomposition. There is only one exception in Table~\ref{tab:manymodalqa_ablation}. This is because ManyModalQA has many single-hop queries, so decomposing everything may raises extra noise.   

\textbf{Without Aggregator: }
In this experiment, the final answer aggregator is disabled. While the system still decomposes the query and executes sub-queries iteratively, the final output is taken directly from the last sub-query’s answer, rather than being aggregating from all intermediate results.
In all cases the result quality drops significantly, especially F1-score score and Semantic Hit. This is because our aggregator comprehend the sub-queries results and ensure the final result is concise and more accurate. Hit score may be better without aggregator, but this is simply because the system tends to output long result, which has a larger chance to hit the ground-truth keywords but is actually not a correct answer.  



\textbf{Single Adapter (LOTUS): }
We also evaluate the impact of the number of adapters. In the Single Adapter (LOTUS) configuration (the last row in the three tables), the system uses only the LOTUS adapter and disables the router as there is only one adapter to route. All other components from checker to aggregator are used as normal. 
Comparing to the full versions of \ours{}, single adapter system shows a lower performance, but it still outperforms the corresponding baseline method. Specifically, single LOTUS adapter \ours{} achieves much higher F1-score (39.76 vs 9.47) and Semantic Hit (57.14 vs 53.90) on MultiModalQA, and all three metrics are significantly better than the LOTUS baseline on M2QA: F1-score 46.58 vs 5.43, Hit 69.06 vs 46.11, Semantic Hit 65.27 vs 38.72. These mean that (1) more adapters do increase the overall performance of \ours{}, proving that \ours{} effectively combines the specialities from different semantic systems; (2) The architecture and workflow design of \ours{} effectively enhances the query processing quality of the underlying semantic systems.

\begin{table}[ht]
\centering
\small
\caption{Average Query Runtime comparison (in seconds)}
\vspace{-10pt}
\label{tab:runtime_all_ablations}
\begin{tabular}{l|cccc}
\toprule
Method
& ManyModalQA
& MultiModalQA
& M2QA 
& Text2Vis\\
\midrule

Lotus                & 6.63 & 7.34 & 5.75 & 10.72 \\
Single-Model         & 6.96 & 7.76 & 6.61 & 9.62 \\
LlamaIndex           & 3.98 & 5.72 & 4.46 & 10.68 \\
DSPy                 & 7.13 & 8.68 & 11.15 & 13.27 \\

\midrule
\multicolumn{5}{l}{\textbf{Meta-Engine (MLRouter)}} \\
\midrule
Full              & 7.76  & 10.59 & 15.91 & 18.06 \\
W/O Checker           & 10.76 & 14.54 & 18.29 & 21.80 \\

\midrule
\multicolumn{5}{l}{\textbf{Meta-Engine (StatRouter)}} \\
\midrule
Full              & 8.28  & 10.75 & 20.81 & 22.66 \\
W/O Checker           & 11.67 & 16.11 & 23.70 & 24.86 \\

\bottomrule
\end{tabular}
\vspace{-15pt}
\end{table}


\subsection{Efficiency Analysis}
\label{sec:efficiency-exp}
We measure end-to-end efficiency on several datasets and report the results in Table~\ref{tab:runtime_all_ablations}. 
Particularly, we include the results of \ours{} without complexity checker, to show the importance of dynamic query decomposition to the overall efficiency. 
On ManyModalQA and MultiModalQA, the full version of \ours{} achieves competitive runtime to several baselines. For example, \ours{} (MLRouter) runs in 7.76s on ManyModalQA, close to DSPy (7.13s) and Single-Model (6.96s); on MultiModalQA, \ours{} (MLRouter) takes 10.59s, competitive to DSPy at 8.68s. On M2QA and Text2Vis, \ours{} incurs higher latency (which is still acceptable like 15.91 VS 11.15 of DSPy), primarily due to the increased number of operator generation iterations required by the more complex queries in these datasets. Specifically, the operator generation stage averages 1.05–1.13 iterations per query on MultiModalQA, while it averages 1.16–1.28 on ManyModalQA, and 1.52–1.81 on M2QA, showing a clear positive correlation between iteration count and runtime. This suggests that improving the accuracy of the operator ranking mechanism is a promising direction for reducing latency. 

As discussed in Section~\ref{sec:ablation-exp}, removing the checker consistently increases runtime across all datasets (e.g., from 7.76s to 10.76s on ManyModalQA with MLRouter, and from 10.59s to 14.54s on MultiModalQA), as the maximum number of subqueries have to be executed for each query. This further confirms the importance and effectiveness of the dynamic query decomposition in both accuracy and efficiency.
Additionally, the pipeline with statistic router is slower than that with learned router. This shows another advantage of learned router: it has a more accurate favor for the efficient adapter while pursuing the best quality.

\section{Conclusion}

In this paper, we presented Meta Engine, a unified framework for building semantic query pipelines that coordinates multiple LLM-based query systems to support complex multimodal tasks. Meta Engine provides easy-to-use APIs and flexible and composable components (from query decomposer to result aggregator) to minimize the users' cost to customize their pipelines and integrate multiple LLM systems. 
Extensive experimental results across diverse multimodal datasets demonstrate that the pipelines built on Meta Engine achieve superior query processing quality while maintaining competitive efficiency comparing to state-of-the-art LLM-based query systems.

\section{Self-Declaration of Related Submissions}
We confirm that there are no related submissions by the same authors currently under review at PVLDB or any other venue.

\bibliographystyle{ACM-Reference-Format}
\bibliography{main}

\end{document}